\documentclass[12pt]{article}
\usepackage{a4wide,latexsym,graphicx,epsfig,psfrag,here}
\usepackage{amsmath} 
\usepackage{longtable}
\usepackage{amssymb} 
\usepackage{bbm}
\usepackage{cite}

\newcommand{\cL}{{\cal L}}
\newcommand{\dg}{\dagger}
\newcommand{\nn}{\nonumber \\}
\newcommand{\no}{\nonumber}

\newcommand{\ve}{\varepsilon}

\newcommand{\DAl}{\raise-.5mm\hbox{\Large$\Box$}}

\begin{document}
\parskip=3pt plus 1pt

\begin{titlepage}
\vskip 1cm
\begin{flushright} 
{
UWThPh-2007-10 \\
May 2007
}
\end{flushright}
\vskip 1cm

\setcounter{footnote}{0}
\renewcommand{\thefootnote}{\fnsymbol{footnote}}

\begin{center} 
{\LARGE \bf Tensor meson exchange  at low energies\footnote{Work 
supported in part by EU contracts HPRN-CT2002-00311 (EURIDICE) and
MRTN-CT-2006-035482 (FLAVIAnet)}} 
\\[30pt] 
{\normalsize \bf  
G. Ecker and C. Zauner  } 

\vspace{1cm}

Faculty of Physics, University of Vienna \\ 
Boltzmanngasse 5, A-1090 Vienna, Austria \\[10pt] 

\end{center} 

\setcounter{footnote}{0}
\renewcommand{\thefootnote}{\arabic{footnote}} 

\vfill

\begin{abstract}
\noindent 
We complete the analysis of meson resonance contributions to chiral
low-energy constants of $O(p^4)$ by including all quark-antiquark
bound states with orbital angular momentum $\le 1$. Different tensor
meson Lagrangians used in previous work are shown to produce the same
final results for the low-energy constants once QCD short-distance
constraints are properly implemented. We also discuss the possible
relevance of axial-vector mesons with odd C-parity ($J^{PC}=1^{+-}$).  
\end{abstract}

\vfill

\end{titlepage} 
\newpage
\addtocounter{page}{1}



\section{Introduction} 
\label{sec:intro}
\renewcommand{\theequation}{\arabic{section}.\arabic{equation}}
\setcounter{equation}{0}
Chiral perturbation theory (CHPT) 
\cite{Weinberg:1978kz,Gasser:1983yg,Gasser:1984gg} is the effective 
field theory of the
Standard Model at low energies. Its explicit degrees of freedom are
the pseudoscalar mesons, the pseudo-Goldstone bosons of spontaneous
chiral symmetry breaking. Since CHPT is to describe all manifestations
of the Standard Model at low energies, heavier degrees of
freedom must be present in the theory as well. As in all effective
theories, heavy states manifest themselves in the coupling
constants of the effective theory called low-energy constants (LECs)
in CHPT. Realistic estimates of chiral LECs are
essential for the predictive power of CHPT.

In the strong mesonic sector, both empirical and theoretical evidence
suggest that chiral LECs are saturated by the lowest-lying meson
resonances. In particular, the LECs of next-to-leading order,
$O(p^4)$, are dominated by vector- and axial-vector meson exchange
\cite{Ecker:1988te,Ecker:1989yg,Donoghue:1988ed}, to a lesser extent
by scalar and pseudoscalar exchange (see
Refs.~\cite{Kaiser:2005eu,Ecker:2007dj} for recent reviews). 
A systematic framework 
for incorporating meson resonance exchange is based on the $1/N_c$
expansion of QCD\cite{'tHooft:1973jz,Peris:1998nj,Pich:2002xy}. 
Although large $N_c$ predicts an infinite number of mesons
(stable to leading order in $1/N_c$), it is clear that the
lowest-lying states will be most important. In fact, meson resonance
exchange contributions to the LECs of $O(p^4)$ scale as $c_R/M_R^2$
for a resonance with mass $M_R$ where $c_R$ is a measure of resonance
couplings. Both the strong coupling to pseudoscalars and the
comparatively low masses of the lightest vector meson nonet are 
responsible for the success of vector meson dominance. The relevance 
of other multiplets must be investigated case by case.

Of all $\overline{q} q$ bound states with orbital angular momentum $L \le
1$, only the states with $J^{PC}=2^{++}$ (tensor mesons) and
$J^{PC}=1^{+-}$ (axial-vector mesons with odd C-parity) still need to
be analysed. Although 
tensor meson contributions to chiral LECs were already considered by 
Donoghue et al. \cite{Donoghue:1988ed} nearly 20 years ago, very
different predictions can be found in the literature
\cite{Suzuki:1993zs,Toublan:1995bk,Ananthanarayan:1998hj,
Dobado:2001rv,Katz:2005ir}.
On the other hand, the influence of $1^{+-}$ resonances on chiral LECs
has not been considered previously. It is the purpose of the present
work to settle the issue of tensor meson exchange and to investigate
the possible relevance of the $1^{+-}$ nonet at low energies. We 
work in the framework of chiral $SU(3)$ but compare also with previous
predictions for tensor contributions within chiral $SU(2)$.    

In Sec.~\ref{sec:resex} we recall the phenomenological status of the
$O(p^4)$ LECs $L_1,L_2,L_3,L_9$ and $L_{10}$ and the evidence for
resonance saturation. The importance of incorporating the proper
short-distance constraints is exemplified for the vector form factor
of the pion. In Sec.~\ref{sec:tensor} we introduce the chiral
Lagrangian for tensor mesons and the most general coupling of lowest
order to the pseudoscalar mesons. It is shown that two of the three
couplings do not contribute to tensor meson decay
amplitudes. In the following
section we investigate the constraints from axiomatic field theory for
elastic meson-meson scattering. Applied to the tensor meson exchange
amplitudes, the constraints fix the tensor contributions to the LECs
$L_1$, $L_2$ and $L_3$ uniquely in terms of the single coupling
constant governing tensor meson decays. We compare our results with 
previous work on tensor meson exchange. In
Sec.~\ref{sec:vector} we turn to the axial-vector mesons with
$J^{PC}=1^{+-}$. Although they superficially contribute to the same
LECs as the vector mesons, albeit with opposite sign, the same
short-distance constraints that determine the vector meson
contributions uniquely \cite{Ecker:1989yg} imply the absence of all 
$1^{+-}$ contributions to the LECs of $O(p^4)$. Sec.~\ref{sec:concl}
summarizes our conclusions. Two appendices contain basic features of
the Lagrangians for symmetric (spin 2) and antisymmetric tensor fields
(spin 1).

\section{Low-energy couplings and resonance exchange} 
\label{sec:resex}
\renewcommand{\theequation}{\arabic{section}.\arabic{equation}}
\setcounter{equation}{0}
Besides the leading-order Lagrangian
\begin{equation}
\cL_2 = \displaystyle\frac{F^2}{4} \langle u^\mu u_\mu + \chi_+
\rangle
\label{eq:L2}
\end{equation}
we shall be concerned with the strong chiral Lagrangian of $O(p^4)$
\cite{Gasser:1984gg}. For chiral $SU(3)$, it can be written in the
form  
\begin{eqnarray}
\cL_4 &=& L_1 \langle  u^\mu u_\mu \rangle^2 + L_2 \langle u^\mu u^\nu 
\rangle \langle u_\mu u_\nu \rangle + L_3 \langle ( u^\mu u_\mu )^2 
\rangle \nn
&& + L_4 \langle u^\mu u_\mu \rangle \langle \chi_+ \rangle 
+ L_5 \langle u^\mu u_\mu \chi_+ \rangle 
+ L_6 \langle \chi_+ \rangle^2 + L_7 \langle \chi_- \rangle^2  \nn
&& + \displaystyle\frac{L_8}{2} \langle \chi_+^2 + \chi_-^2 \rangle 
- i L_9 \langle f_{+ \,\mu \nu} u^\mu u^\nu \rangle 
+ \displaystyle\frac{L_{10}}{4} \langle f_{+ \,\mu \nu}f_+^{\mu\nu} 
- f_{- \,\mu \nu}f_-^{\mu\nu} \rangle 
\label{eq:L4}
\end{eqnarray}
in terms of 10 LECs $L_1,\dots,L_{10}$.
The various matrix fields are defined as usual (see, e.g.,
Ref.~\cite{Ecker:1988te}):
\begin{eqnarray}
u_\mu &=& i \{ u^\dg(\partial_\mu - i r_\mu)u - 
u(\partial_\mu - i \ell_\mu) u^\dg\} \nn
\chi_\pm &=&  u^\dg \chi u^\dg \pm u \chi^\dg u \nn
f_{\pm \, \mu\nu} &=& u F_{L \; \mu\nu} u^\dg \pm u^\dg F_{R \;
\mu\nu} u  \nn
F_{R \, \mu\nu} &=& \partial_\mu r_\nu - \partial_\nu r_\mu -
i[r_\mu,r_\nu] \nn
F_{L \; \mu\nu} &=& \partial_\mu \ell_\nu - \partial_\nu \ell_\mu -
i [\ell_\mu,\ell_\nu]~.  
\end{eqnarray} 
Here, $u(\phi)$ is the coset space element parametrized by the
Goldstone fields. The external matrix fields
$v_\mu,a_\mu,s,p$ are contained in  $r_\mu = v_\mu + a_\mu$, 
$\ell_\mu = v_\mu - a_\mu$, $\chi= 2 B (s+ip)$. The symbol
$\langle \dots \rangle$ denotes the 3-dimensional flavour trace. The
LECs of lowest order $B$, $F$ are related to the quark condensate and
to the pion decay constant, respectively.

Only the LECs $L_1$, $L_2$, $L_3$, $L_9$ and $L_{10}$ will be relevant
for the following analysis. The present phenomenological status and the
resonance contributions of the standard variety $V(1^{--})$,
$A(1^{++})$,  $S(0^{++})$ are collected in Table \ref{tab:LECSp4} 
($P(0^{-+})$ exchange does not contribute in
this case). Keeping in mind that resonance exchange does not fix
the renormalization scale of the renormalized LECs $L_i(\mu)$, the
overall agreement with the phenomenological values suggests that $V$,
$A$ and $S$ already saturate the LECs in Table \ref{tab:LECSp4}. In
fact, scalar exchange makes only a relatively small 
contribution to $L_3$. On the other hand, the situation in Table
\ref{tab:LECSp4} certainly leaves room for additional contributions. 
We therefore include all meson resonances corresponding to 
$\overline{q} q$ bound states with orbital angular momentum $L \le
1$. From the point-of-view of quantum numbers, states with
$J^{PC}=2^{++}$ and  $1^{+-}$ could in principle contribute to some of
the LECs in Table \ref{tab:LECSp4}. 
\begin{table}
\centering
\renewcommand{\arraystretch}{1.2}
\begin{tabular}{|c||rrcc|}  \hline 
 & & &  & \\[-6pt]
i & Ref.~\cite{Gasser:1984gg}  \hspace*{.2cm} &
Ref.~\cite{Bijnens:2006zp} \hspace*{.3cm}   &  &
 \mbox{} \hspace*{.1cm}  Ref.~\cite{Ecker:1988te} \hspace*{.1cm}
\\[4pt] 
\hline
 & & & &  \\[-4pt]
  1  & 0.7 $\pm$ 0.3 & 0.43 $\pm$ 0.12 &  & 0.6     \\[2pt]
  2  & 1.3 $\pm$ 0.7 &  0.73 $\pm$ 0.12 &   & 1.2   \\[2pt]
  3  & \hspace*{.5cm} $-$4.4 $\pm$ 2.5 & \hspace*{.7cm} $-$2.35 $\pm$
     0.37 &  & \hspace*{.3cm} $-$3.0 \hspace*{.6cm}  \\[2pt]
  9  & 6.9 $\pm$ 0.7 & 5.93 $\pm$ 0.43 &  &
     6.9 \hspace*{-.15cm}   \\[2pt]
 10  & $-$5.5 $\pm$ 0.7 & $-$5.09 $\pm$ 0.47  &  & $-$6.0 \hspace*{.1cm}
     \\[4pt]
\hline
\end{tabular}

\caption{Phenomenological values and theoretical estimates for the
$SU(3)$ LECs $L_i(M_\rho)$ in units of $10^{-3}$. The first column
  shows the original values of Ref.~\cite{Gasser:1984gg}, the second
  displays the current values taken from Ref.~\cite{Bijnens:2006zp} and
  references therein. The third column contains the
  resonance saturation results of Ref.~\cite{Ecker:1988te}. The 
  value for $L_9$ was taken as input in Ref.~\cite{Ecker:1988te}.}
\label{tab:LECSp4}
\end{table}
 
We follow here the traditional approach of chiral
resonance Lagrangians \cite{Ecker:1988te}. Compared to studying
Green functions directly with a large-$N_c$ inspired ansatz,
the Lagrangian approach offers the possibility  of integrating out 
the resonance fields once and for all in the generating functional of 
Green functions (to leading order in $1/N_c$), thereby
generating all contributions of a given order. In addition,
chiral symmetry is of course guaranteed so that chiral Ward identities
are satisfied automatically.

A priori, the chiral resonance Lagrangian knows nothing about the
short-distance structure of QCD. Therefore, the Lagrangian approach
must always be supplemented by 
short-distance constraints \cite{Ecker:1989yg}. This will turn out to 
be especially important for resonance contributions of the type 
$J^{PC}=2^{++}$ and $1^{+-}$. It will be sufficient to implement the
same constraints that were used to establish the uniqueness of vector
and axial-vector contributions \cite{Ecker:1989yg} to the LECs in 
Table \ref{tab:LECSp4}.

Short-distance constraints refer to Green functions
or amplitudes but not to special resonance exchanges. Is it then 
legitimate to apply those constraints to a given resonance exchange 
contribution if only the sum of (an infinite number of) such exchanges 
must satisfy the constraints? 

An instructive example is provided by the vector form factor of the
pion $F^\pi_V(t)$. From the asymptotic behaviour of the $I=1$ vector
current two-point function in QCD we know \cite{Floratos:1978jb} that
$F^\pi_V(t)$ satisfies a dispersion relation with at most one
subtraction:   
\begin{eqnarray} 
F^\pi_V(t) = 1 + \displaystyle\frac{t}{\pi} \displaystyle\int_0^\infty
dt^\prime \displaystyle\frac{{\rm Im}\,F^\pi_V(t^\prime)}{t^\prime 
(t^\prime  - t - i\epsilon)} ~. 
\label{eq:FVsub}
\end{eqnarray}
To first nontrivial order in the low-energy expansion, $F^\pi_V(t)$ is
given by \cite{Gasser:1984ux}
\begin{equation}
F^\pi_V(t) = 1 + \displaystyle\frac{2}{F^2} L_9(\mu) \,t +
\displaystyle\frac{2}{F^2} \Phi(t,M_\pi^2,M_K^2;\mu) + O(p^6) ~, 
\end{equation}
where the function $\Phi(t,M_\pi^2,M_K^2;\mu)$ accounts for pion and
kaon loops. The slope of the form factor gives rise to the sum rule
\begin{equation}
L_9(\mu) + \displaystyle\frac{d \Phi}{dt} (0,M_\pi^2,M_K^2;\mu)  +
O(p^6) = \displaystyle\frac{F^2}{2\pi} \displaystyle\int_0^\infty dt
\;\displaystyle\frac{{\rm Im} F^\pi_V(t)}{t^2} ~.    
\label{eq:L9sr}
\end{equation}
Both the scale dependence of $L_9$ and the loop function $\Phi$ are
non-leading in $1/N_c$. Since LECs do not depend on light quark
masses, we can take the chiral limit to eliminate contributions
from higher-order LECs. At leading order in $1/N_c$, the absorptive
part is given by
\begin{equation}
{\rm Im} F^\pi_V(t) = \displaystyle\frac{2\pi}{F^2} \displaystyle\sum_R
\kappa_R M_R^2\, \delta(t - M_R^2) 
\end{equation}
giving rise to the form factor
\begin{equation}
F^\pi_V(t) = 1 + \displaystyle\frac{2 t}{F^2} \displaystyle\sum_R
\displaystyle\frac{\kappa_R}{M_R^2 - t - i\epsilon} ~.
\end{equation} 
To leading order in $1/N_c$, $L_9$ is therefore of the familiar form
\begin{equation}
L_9 = \displaystyle\sum_R \displaystyle\frac{\kappa_R}{M_R^2} 
\end{equation}
where $\kappa_R$ is related to the product of resonance couplings
to the electromagnetic current and to two pions. To a given resonance
we can associate unambiguously the contribution
\begin{equation}
L_9^R = \displaystyle\frac{\kappa_R}{M_R^2} 
\end{equation}
even though only the total $L_9$ emerges from the sum rule
(\ref{eq:L9sr}). Of course, the same conclusion is reached by
subjecting each resonance separately to the short-distance
constraints, which in the present case are encoded in the
once-subtracted dispersion relation (\ref{eq:FVsub}). 

In the approach with chiral resonance Lagrangians, consistency with
short-distance constraints is not automatic. In general, local
contributions from the chiral Lagrangian (\ref{eq:L4}) of $O(p^4)$
must be added to achieve consistency \cite{Ecker:1989yg}. An 
important lesson can be drawn
from the example of the pion form factor: only pole terms in the
form factor contribute to the LEC $L_9$. This will be of special
relevance for the evaluation of $1^{+-}$ contributions to the LECs in
Sec.~\ref{sec:vector} .

\section{Tensor meson exchange} 
\label{sec:tensor}
\renewcommand{\theequation}{\arabic{section}.\arabic{equation}}
\setcounter{equation}{0}

In this section we compute the effective action due to the exchange of
the lowest-lying nonet of tensor mesons with $J^{PC}=2^{++}$. 
We describe these particles by a symmetric hermitian rank-2 tensor
field 
\begin{eqnarray}
 T_{\mu\nu} = T_{\mu\nu}^0 \frac{\lambda_0}{\sqrt{2}}+ 
\frac{1}{\sqrt{2}} \sum_{i=1}^8 \lambda_i T_{\mu\nu}^{8,i}
	& , 	& T_{\mu\nu} = T_{\nu\mu}~. 
\end{eqnarray}
The octet and the singlet components are given by
\begin{eqnarray} 
\frac{1}{\sqrt{2}} \sum_{i=1}^8 \lambda_i T^{8,i} = \left(
\begin{array}{ccc}
   \frac{a_2^0}{\sqrt{2}}+\frac{f_2^8}{\sqrt{6}} & a_2^+ & K_2^{*+} \\
   a_2^- &  -\frac{a_2^0}{\sqrt{2}}+\frac{f_2^8}{\sqrt{6}} & K_2^{*0} \\
	K_2^{*-} & \bar K_2^{*0} & 	-\frac{2 f_2^8}{\sqrt{6}}
	\end{array}
	\right) &,& T^0 = f_{2}^0 ~.
\end{eqnarray}  
The tensor nonet couples to pseudoscalar mesons via the Lagrangian
(see App.~\ref{app:tensor})
\begin{eqnarray}
	\mathcal L = - \frac{1}{2} \left\langle T_{\mu\nu} 
D^{\mu\nu,\rho\sigma}_T  T_{\rho\sigma} \right\rangle + 
\left\langle T_{\mu\nu} J^{\mu\nu}_T \right\rangle 
\label{eq:L_Tnonet}
\end{eqnarray}
with a symmetric tensor current $J^{\mu\nu}_T=J^{\nu\mu}_T$.
For the octet part, the derivatives in $D^{\mu\nu,\rho\sigma}_T$ in 
(\ref{eq:D_T}) must be replaced by chirally covariant derivatives.
This modification will not affect the structure of the
effective action up to $O(p^4)$.

The most general symmetric tensor current $J^{\mu\nu}_T$ of $O(p^2)$
(relevant for LECs of $O(p^4)$) consists of three terms
\cite{Bellucci:1994eb}:
\begin{eqnarray}
J^{\mu\nu}_T &=&  J_1^{\mu\nu} + g^{\mu\nu} J_2    \nn
J_1^{\mu\nu} = g_T \left\{u^\mu,u^\nu
\right\}~, &&  J_2 = \beta u^\mu u_\mu + \gamma~\chi_+ ~.
\label{eq:gT}
\end{eqnarray}
In the following section, we will consider elastic meson-meson
scattering. For this purpose, we can use the free tensor propagator
(\ref{eq:tensorprop}) in the effective action for meson-meson
scattering:
\begin{eqnarray}
S_T^{\rm eff}(M M \to M M) = \displaystyle\frac{1}{2} 
\displaystyle\int d^4x \;d^4y
\left\langle J^{\mu\nu}_T(x) G^T_{\mu\nu,\rho\sigma}(x-y)
J^{\rho\sigma}_T(y) \right\rangle ~. 
\end{eqnarray}
It will be convenient to separate the contributions of $J_1^{\mu\nu}$
and $J_2$ to the effective action. Due to the structure of the
propagator (\ref{eq:tensorprop}), the effective action takes the form
\begin{eqnarray}
S_T^{\rm eff}(M M \to M M) &=& \displaystyle\frac{1}{2} 
\displaystyle\int d^4x \;d^4y
\left\langle J^{\mu\nu}_1(x) G^T_{\mu\nu,\rho\sigma}(x-y)
J^{\rho\sigma}_1(y) \right\rangle 
\label{eq:J1} \\
&& \hspace*{-4.5cm} - \displaystyle\frac{1}{3 M_T^2} \displaystyle\int d^4x 
\left\langle  J_2(x) \left(g_{\mu\nu} - 2 M_T^{-2} \partial_\mu
\partial_\nu  \right) J^{\mu\nu}_1(x)\right\rangle - 
\displaystyle\frac{1}{3 M_T^2} \displaystyle\int d^4x \left\langle 
J_2(x) \left(2 -  M_T^{-2} \DAl \right) J_2(x) \right\rangle . \no 
\end{eqnarray}
The important observation is that the tensor current $g^{\mu\nu} J_2$
contributes only to the local part of the effective action of $O(p^4)$
and higher. As we shall see in the following section,
such local actions must in fact be added to the bare tensor exchange
in order to satisfy appropriate short-distance
constraints. Thus, the couplings $\beta,\gamma$ in the current
$J_2$ can always be absorbed in the effective chiral Lagrangians. We 
therefore set them to zero in this section without loss of
generality. Nevertheless, it will turn out to be convenient to 
reinstall $\beta \neq 0$ for the discussion of short-distance constraints 
in pion-pion scattering in the next section. 

The couplings $\beta$ and $\gamma$ are arbitrary because,
in contrast to the coupling constant $g_T$ in
(\ref{eq:gT}), they cannot be determined from partial decay
widths of tensor resonances. 
The tensor field is traceless on-shell ($\ve^\mu_{\,\mu}=0$ in
(\ref{eq:ME_T})) so that $\beta$ and $\gamma$ do not enter matrix 
elements for tensor meson decays.

In order to determine the LECs of $O(p^4)$ due to tensor exchange, we
need to take the leading term of the propagator (\ref{eq:tensorprop})
in an expansion in $1/M_T^2$:
\begin{equation}
G^T_{\mu\nu,\rho\sigma}(x)|_{O(M_T^{-2})} = \displaystyle\frac{1}{6 M_T^2}
\left\{ 3\left(g_{\mu\rho} g_{\nu\sigma} + g_{\mu\sigma} g_{\nu\rho}
\right) - 2 g_{\mu\nu} g_{\rho\sigma} \right\} \delta^{(4)}(x)~.
\end{equation} 
From the first term in the effective action (\ref{eq:J1}) we then obtain
an effective Lagrangian $\cL_{4,{\rm bare}}^T$ of $O(p^4)$ from tensor
exchange: 
\begin{eqnarray} 
\cL_{4,{\rm bare}}^T = \displaystyle\frac{g_T^2}{2 M_T^2} \left\{
\left\langle u_\mu u^\mu \right\rangle^2 + 2 \left\langle u_\mu u_\nu
\right\rangle  \left\langle u^\mu u^\nu \right\rangle - 
\displaystyle\frac{10}{3} \left\langle u^\mu u_\mu u^\nu u_\nu
\right\rangle \right\} ~.
\label{eq:LTbare}
\end{eqnarray}   
Comparing with the general Lagrangian (\ref{eq:L4}) of $O(p^4)$, we
find the following (bare) LECs due to tensor exchange:
\begin{eqnarray}
L_{1,{\rm bare}}^T = \displaystyle\frac{g_T^2}{2 M_T^2}~, \qquad
L_{2,{\rm bare}}^T = 2 L_{1,{\rm bare}}^T ~, \qquad
L_{3,{\rm bare}}^T = - \displaystyle\frac{5 g_T^2}{3 M_T^2} ~. 
\label{eq:LiTbare}
\end{eqnarray} 
We denote the Lagrangian (\ref{eq:LTbare}) and the LECs
(\ref{eq:LiTbare}) as bare quantities because we still have to check
for consistency with the short-distance structure of QCD.
As we shall see in the next section,the short-distance constraints will
modify the bare LECs substantially. Finally, we note that
exchange of the tensor nonet is of course compatible with the
large-$N_c$ prediction $L_2 = 2 L_1$.

\section{Short-distance constraints for tensor exchange} 
\label{sec:SDtensor}
\renewcommand{\theequation}{\arabic{section}.\arabic{equation}}
\setcounter{equation}{0}
The LECs $L_1, L_2, L_3$ all contribute to meson-meson scattering. It
will turn out to be sufficient to investigate 
forward dispersion relations for elastic meson-meson scattering
amplitudes. Since the $L_i$ do not depend on light quark masses, all
calculations will be performed in the chiral limit. 

We briefly recall the well-known structure of the forward dispersion
relation for an elastic channel with $s \leftrightarrow u$ symmetry,
e.g., $\pi^+ \pi^0 \to \pi^+ \pi^0$ (for a recent account see 
Ref.~\cite{Ecker:2007dj}). In this case, general quantum field theory
guarantees \cite{Froissart:1961ux,JM64} that the scattering amplitude
$A(\nu,t)$ satisfies a once-subtracted forward dispersion relation in
$\nu=(s-u)/2$: 
\begin{equation} 
A(\nu,t=0) = A(0,0) + \displaystyle\frac{\nu^2}{\pi}
\displaystyle\int_0^\infty d\nu^{\prime\,2}
\displaystyle\frac{Abs~A(\nu^\prime,0)}{\nu^{\prime\,2} \left(\nu^2 -
\nu^{\prime\,2} \right)}   ~.
\label{eq:disp}
\end{equation}
In the chiral limit and to leading order in $1/N_c$, exchange of a 
resonance gives rise to an amplitude
\begin{equation} 
A(\nu,0) = \displaystyle\frac{c_R \nu^2}{\nu^2- M_R^4} 
\end{equation} 
where $c_R$ is related to the partial decay width
$\Gamma(R \to M M)$ in this case. On the other hand, resonance
exchange from  a chiral resonance Lagrangian such as
(\ref{eq:L_Tnonet}) will produce an amplitude of the general form
\begin{equation} 
A_R (\nu,0) = \displaystyle\frac{P_R(\nu^2)}{\nu^2- M_R^4} ~,
\end{equation}
with a polynomial $P_R(\nu^2)$ satisfying the on-shell condition 
$P_R(M_R^4) = c_R M_R^4$. Decomposing the polynomial $P_R(\nu^2)$ as
\begin{equation}
P_R(\nu^2)=P_R(M_R^4)+\left(\nu^2- M_R^4
\right) \overline{P_R}(\nu^2) ~,
\end{equation}
the equality $A_R (\nu,0) = A(\nu,0)$ forces $\overline{P_R}(\nu^2)$
to be a constant,
\begin{equation} 
\overline{P_R}(\nu^2) = c_R ~.
\end{equation}
This will not be the case for our tensor meson Lagrangian
(\ref{eq:L_Tnonet}). Therefore, the dispersion
relation (\ref{eq:disp}) requires the addition of a 
(counterterm) polynomial $P_c(\nu^2)$ from the effective chiral
Lagrangians of $O(p^4)$ and higher: 
\begin{equation}
A_R (\nu,0) = P_c(\nu^2) + \overline{P_R}(\nu^2) + 
  \displaystyle\frac{P_R(M_R^4)}{\nu^2- M_R^4} ~.
\end{equation}
The counterterm polynomial $P_c(\nu^2)$ is then fixed by the
short-distance  constraint to satisfy
\begin{equation}
P_c(\nu^2) + \overline{P_R}(\nu^2) = c_R ~,
\end{equation}
ensuring at the same time the correct low-energy behaviour of the 
resonance exchange amplitude:
\begin{equation}
A_R (\nu,0) = A(\nu,0) = - \displaystyle\frac{c_R}{M_R^4} \nu^2
+ O(p^8) ~. 
\end{equation}
The coefficient of $\nu^2$ depends only on the mass and on the 
partial decay width of the resonance and it defines the resonance
contribution  to a certain combination of the $L_i$.

Even though we are interested in the low-energy behaviour, 
the same conclusion is obtained by comparing the high-energy
behaviour of $A_R (\nu,0)$ with $A(\nu,0)$ :
\begin{eqnarray}
\displaystyle\lim_{\nu^2 \to \infty} A(\nu,0) = c_R =
\displaystyle\lim_{\nu^2 \to \infty} A_R(\nu,0) = 
\displaystyle\lim_{\nu^2 \to \infty} \left( P_c(\nu^2) +
\overline{P_R}(\nu^2)\right) ~.
\end{eqnarray}
It will often be more convenient to investigate the high-energy
behaviour.

\subsection{Elastic meson-meson scattering}

The meson-meson scattering amplitude due to tensor meson exchange can
be extracted from the effective action (\ref{eq:J1}). Following the
discussion in Sec.~\ref{sec:tensor}, we are led to include only the
interaction term $J_1^{\mu\nu}$. It turns out that the short-distance
constraints embodied in the forward dispersion relation
(\ref{eq:disp}) would then require the addition of local
terms not only of $O(p^4)$ but also of $O(p^6)$. 

By a judicious choice of the (a priori) arbitrary coupling constant
$\beta$ in (\ref{eq:gT}) we can avoid having to include terms of $O(p^6)$
at this stage where we are only interested in the LECs of
$O(p^4)$. The specific value of $\beta$ ensuring the absence of $p^6$
terms is
\begin{equation}
\beta = - g_T 
\label{eq:bspecial}
\end{equation}
corresponding to a special structure of the tensor coupling 
$J_T^{\mu\nu}$. In this case, the bilinear terms in $u_\mu$ in
$J^{\mu\nu}_T$ occur in the same combination as in the energy-momentum 
tensor \cite{Donoghue:1991qv} associated with the
lowest-order chiral Lagrangian (\ref{eq:L2}). It was already observed
by Bellucci et al. \cite{Bellucci:1994eb} that this choice of
$J_T^{\mu\nu}$ leads to a smoother high-energy behaviour than in the
general case.

We hasten to emphasize that our final values for the LECs $L_1, L_2,
L_3$ will be completely independent of the choice of $\beta$. As 
already pointed out, $\beta$ appears in
the scattering amplitude only through polynomial terms 
that can always be absorbed in contributions from the chiral
Lagrangians of $O(p^4)$ (and $O(p^6)$ in general). The main advantage
of the choice (\ref{eq:bspecial}) is that it allows us to omit the
qualifying statement ``up to terms of $O(p^6)$'' after every other
equation.
 
We now consider pion-pion scattering. The scattering amplitude
$ T(\pi_a \pi_b \rightarrow \pi_c \pi_d) \equiv T_{ab,cd}(s,t,u)$ can 
be expressed in terms of a single function $A(s,t,u)=A(s,u,t)$ as 
\begin{equation}
T_{ab,cd}(s,t,u) = A(s,t,u) \;\delta_{ab} \delta_{cd} + A(t,s,u)
\;\delta_{ac} \delta_{bd} + A(u,t,s) \;\delta_{ad} \delta_{bc}~.
\end{equation}
From the effective action (\ref{eq:J1}) we obtain the tensor exchange
amplitude in the chiral limit (for $\beta = - g_T$):
\begin{equation}
A_T(s,t,u) = \displaystyle\frac{2 g_T^2}{F^4 (M_T^2 - s)} \left[(t-u)^2 -
\displaystyle\frac{s^2}{3} \right] ~. 
\end{equation}
In order to satisfy the short-distance constraints, we have to
add an explicit local amplitude from the $O(p^4)$ Lagrangian (\ref{eq:L4}):
\begin{equation} 
A_{\rm SD}(s,t,u) = \displaystyle\frac{4}{F^4} \left[(2 L_1^{\rm SD} +
  L_3^{\rm SD}) s^2 +  L_2^{\rm SD}(t^2+u^2) \right]~. 
\end{equation}
The $L_i^{\rm SD}$ will be determined from the short-distance
constraints but they are of course not the final values of the LECs
associated with tensor meson exchange. The final values are obtained
by expanding the complete amplitude $A_T(s,t,u) + A_{\rm SD}(s,t,u)$
to $O(p^4)$: 
\begin{eqnarray} 
A_T(s,t,u) + A_{\rm SD}(s,t,u) && \nn
&& \hspace*{-4.5cm} =  \displaystyle\frac{2 g_T^2}{F^4 M_T^2} 
\left[(t-u)^2 - \displaystyle\frac{s^2}{3} \right] +
\displaystyle\frac{4}{F^4} \left[(2 L_1^{\rm SD} +
  L_3^{\rm SD}) s^2 +  L_2^{\rm SD}(t^2+u^2) \right] + O(p^6) \nn
&& \hspace*{-4.5cm} = \displaystyle\frac{4}{F^4} \left[s^2 
\left(2 L_1^{\rm SD} +
  L_3^{\rm SD}- \displaystyle\frac{2 g_T^2}{3 M_T^2} \right)
  + (t^2 + u^2) \left( L_2^{\rm SD} + 
\displaystyle\frac{g_T^2}{M_T^2} \right) \right]+ O(p^6).
\end{eqnarray} 
We can immediately read off the total tensor exchange contributions 
$2 L_1^{T} + L_3^{T}$ and $L_2^{T}$ from the last expansion. To obtain
$L_1^T$ and $L_3^T$ separately, we either need another
independent channel, e.g., elastic $K \pi$ scattering, or we appeal to
the large-$N_c$ relation $L_2 = 2 L_1$ that is of course respected by
exchange of a tensor nonet. Both approaches
lead to the same results:
\begin{eqnarray}
L_2^T = 2 L_1^T &=& \displaystyle\frac{g_T^2}{M_T^2} +  L_2^{\rm
  SD} \nn
L_3^T  &=& - \displaystyle\frac{5 g_T^2}{3 M_T^2} +  L_3^{\rm SD}
  ~. \label{eq:LiT}
\end{eqnarray}
Referring back to Eq.~(\ref{eq:LiTbare}), we observe that the bare
tensor contributions to the $L_i$ are identical. This equality is to
some extent accidental because it happens to hold specifically for the
special cases $\beta=0$ (adopted in Sec.~\ref{sec:tensor}) and
$\beta=-g_T$ assumed here. For other values of $\beta$ the bare term 
$L_{3,{\rm bare}}^T$ will in general be different while 
$L_{1,{\rm bare}}^T$, $L_{2,{\rm bare}}^T$ remain unchanged
\cite{diplom}. However, as the following arguments will show, the
total values $L_i^T$ will always be the same.

In order to determine the short-distance induced contributions
$L_i^{\rm SD}$, we consider the following two channels with $s
\leftrightarrow u$ symmetry:
\begin{eqnarray}
A (\pi^+ \pi^0 \rightarrow \pi^+ \pi^0 ) & = & A(t,s,u)
	\label{Tplusnull} \\ 
A (\pi^0 \pi^0 \rightarrow \pi^0 \pi^0 ) & = & A(s,t,u) + A(t,s,u) +
A(u,t,s)~. \label{Tnullnull}
\end{eqnarray} 
The forward scattering amplitude for the $\pi^+ \pi^0$-channel is
therefore 
\begin{equation}
A_T(\nu,0)|_{\pi^+ \pi^0 \rightarrow \pi^+ \pi^0} = 
\displaystyle\frac{8 g_T^2}{F^4 M_T^2} \nu^2 +
\displaystyle\frac{8}{F^4} L_2^{\rm SD} \nu^2 ~. 
\end{equation}
From the general discussion of the forward amplitude at the beginning
of this section we conclude that $A_T(\nu,0)$ must vanish for the case
of $\pi^+ \pi^0$ scattering (absence of a pole term). This constraint 
fixes $L_2^{\rm SD}$ to be
\begin{equation}
 L_2^{\rm SD} = - \displaystyle\frac{g_T^2}{M_T^2}~.
\label{eq:L2SD} 
\end{equation}
For the second channel we find
\begin{equation}
A_T(\nu,0)|_{\pi^0 \pi^0 \rightarrow \pi^0 \pi^0} =
\displaystyle\frac{8 g_T^2 M_T^2}{3 F^4}\displaystyle\frac{\nu^2}{M_T^4 -
  \nu^2} +  \displaystyle\frac{8 g_T^2}{F^4 M_T^2} \nu^2 +
 \displaystyle\frac{8}{F^4} \left(2 L_1^{\rm SD} + 2 L_2^{\rm SD} 
  + L_3^{\rm SD} \right) \nu^2 ~. 
\end{equation}
Together with (\ref{eq:L2SD}), the structure of the forward dispersion
relation requires
\begin{equation}
 2 L_1^{\rm SD} + L_3^{\rm SD} = \displaystyle\frac{g_T^2}{M_T^2}~.
\label{eq:L13SD} 
\end{equation}
As before, we can either appeal to large $N_c$ or investigate 
additional meson-meson scattering channels to arrive at the following 
results for the $L_i^{\rm SD}$:
\begin{eqnarray}
L_2^{\rm SD} = 2 L_1^{\rm SD} = - \displaystyle\frac{g_T^2}{M_T^2}~, 
\qquad L_3^{\rm SD} = \displaystyle\frac{2 g_T^2}{M_T^2}~. 
\end{eqnarray}
In fact, all channels are compatible with these values. 
Inserting into Eqs.~(\ref{eq:LiT}), we obtain the complete
LECs $L_i^T$ due to tensor meson exchange:
\begin{eqnarray}
L_1^T = L_2^T = 0~,  \qquad L_3^T = \displaystyle\frac{g_T^2}{3
  M_T^2}~.
\label{eq:Lifinal} 
\end{eqnarray}
Comparing with the bare LECs (\ref{eq:LiTbare}),
we observe that the short-distance constraints have eliminated
$L_1$ and $L_2$ altogether. Moreover, the absolute magnitude of
$L_3$ is reduced by a factor of five. Once again, we stress that the
so-called bare values (\ref{eq:LiTbare}) have no intrinsic
meaning. Only the final values (\ref{eq:Lifinal}) can be associated with
tensor meson exchange.

\subsection{Numerical discussion and comparison with previous work}
\label{subsec:num}
The tensor coupling constant $g_T$ defined in Eq.~(\ref{eq:gT}) can be
determined from the decay rate $\Gamma (f_2(1270) \rightarrow \pi
\pi)$. To a good approximation (see, e.g.,
Ref.~\cite{Cirigliano:2003yq}), the $f_2(1270)$ is the non-strange
partner of an ideal mixture of the $SU(3)$ singlet and octet
isosinglet states:
\begin{eqnarray}
f_2(1270)_{\mu\nu} = \left(\sqrt{2} T^0_{\mu\nu} + T^{8,8}_{\mu\nu}  
\right)/\sqrt{3} ~.
\end{eqnarray}
To the accuracy needed for our purposes, the assumption of ideal
mixing is completely sufficient.

The decay rate $\Gamma (f_2(1270) \rightarrow \pi\pi)$ is then given
by 
\begin{eqnarray}
\Gamma (f_2(1270) \rightarrow \pi\pi) = 
\displaystyle\frac{g_T^2 M_T^3}{40 \pi F_\pi^4} 
(1 - 4 M_\pi^2/M_T^2)^{5/2}~.
\end{eqnarray}
With $M_T=M(f_2(1270))$ and $\Gamma (f_2(1270) \rightarrow \pi\pi)$
taken from PDG 2006 \cite{Yao:2006px} and with $F_\pi=92.4$ MeV, one
finds 
\begin{equation}
|g_T| = 28 ~{\rm MeV}~.
\end{equation}
This value should be compared with the corresponding vector and scalar
couplings $G_V$ \cite{Ecker:1989yg} and $c_d$
\cite{Moussallam:1994at,Golterman:1999au,Pich:2002xy,Jamin:2001zq} :
\begin{eqnarray}
|G_V| \simeq \displaystyle\frac{F_\pi}{\sqrt{2}}= 65 ~{\rm MeV}, 
\qquad & \qquad 
46 ~\mbox{MeV} = \displaystyle\frac{F_\pi}{2} \lesssim |c_d| 
\lesssim \displaystyle\frac{F_\pi}{\sqrt{2}}~. 
\end{eqnarray}
Thus, the tensor coupling to pions is not much smaller than
the corresponding vector and scalar couplings. Nevertheless, the only
non-zero contribution of tensor exchange to the LECs of $O(p^4)$,
\begin{eqnarray} 
L_3^T = \displaystyle\frac{g_T^2}{3 M_T^2} = 0.16 \cdot 10^{-3}~,
\end{eqnarray} 
is considerably smaller than the sum of vector and scalar
contributions. This is only partly due to the larger mass $M_T$. We
recall that the so-called bare value in Eq.~(\ref{eq:LiTbare}),
\begin{eqnarray}
L_{3,{\rm bare}}^T = - \displaystyle\frac{5 g_T^2}{3 M_T^2} = - 0.80
\cdot 10^{-3}~,  
\end{eqnarray}
would amount to a non-negligible contribution to $L_3$ (see
Table \ref{tab:LECSp4}).

In much of the previous literature, tensor meson exchange was
considered in the framework of chiral $SU(2)$. To $O(p^4)$, the
$SU(3)$ results can be translated to the $SU(2)$ LECs $l_i^T$ through 
the relations \cite{Gasser:1984gg}
\begin{eqnarray}
l_1^T &=& 4 L_1^T + 2 L_3^T \\
l_2^T &=& 4 L_2^T ~.
\end{eqnarray}
The numerical values for $l_1^T,l_2^T$ from different sources are
collected in Table \ref{tab:LECfromLit}. As far as we are aware,
the first determination of tensor contributions to the $l_i$ was
performed by Donoghue et al. \cite{Donoghue:1988ed}. Their results are 
identical to those in Ref.~\cite{Dobado:2001rv} and they correspond 
exactly to our bare LECs in Eq.~(\ref{eq:LiTbare}) ($\beta=0$ in our
notation). Different tensor meson couplings were used in
Refs.~\cite{Suzuki:1993zs,Katz:2005ir}. In the Lagrangian of 
Ref.~\cite{Katz:2005ir}, the $f_2(1270)$ is assumed to couple like the
graviton to the energy-momentum tensor ($\beta= - g_T$).
Of all the previous work, only Toublan \cite{Toublan:1995bk} and
Ananthanarayan \cite{Ananthanarayan:1998hj} took short-distance
constraints into account. In Ref.~\cite{Ananthanarayan:1998hj}
different versions of dispersion
relations for $\pi\pi$ scattering\footnote{Starting at $O(p^6)$,
crossing symmetry imposes additional constraints on resonance exchange
contributions with spin $\ge 2$ \cite{Ananthanarayan:1998hj,
Ananthanarayan:2000ht}.} were analysed to
determine the $f_2$ contribution to the $l_i$.  
Although Toublan used a different Lagrangian for the tensor 
fields and applied slightly different short-distance arguments, we
agree with his results in the $SU(2)$ limit. The
agreement with Refs.~\cite{Toublan:1995bk,Ananthanarayan:1998hj} 
underscores our claim
that the final results for tensor meson exchange to the LECs are model
independent once the high-energy conditions are properly implemented.
On the other hand, the results in Table
\ref{tab:LECfromLit} document rather convincingly that the
high-energy constraints are essential to arrive at unique values for
the contributions of tensor meson exchange.

\begin{table}
\begin{center}
\renewcommand{\arraystretch}{1.2}
\begin{tabular}{|l|c|c|}
	\hline
& &   \\[-6pt]
  &  \mbox{} \hspace*{.5cm}  $l_1^T \cdot 10^{3}$  \mbox{}
 \hspace*{.5cm} &  \mbox{} \hspace*{.5cm}  $l_2^T \cdot 10^{3}$  
\mbox{}  \hspace*{.5cm} \\[4pt]
	\hline
 & &    \\[-8pt]
 Donoghue, Ramirez, Valencia \cite{Donoghue:1988ed} \hspace*{.5cm}  
 & $- 0.6$ & $1.9$ \\[4pt]
 Dobado, Pelaez \cite{Dobado:2001rv} & $- 0.6$ & $1.9$ \\[4pt]
  \hline 
 & &    \\[-8pt]
 Suzuki \cite{Suzuki:1993zs} & $- 0.5$ & $ 2.0$  \\[4pt] 
  \hline 
 & &    \\[-8pt]
 Katz, Lewandowski, Schwartz \cite{Katz:2005ir}  & $- 0.7$ & $2.1$  
 \\[4pt]
  \hline
 & &    \\[-8pt]
 Toublan \cite{Toublan:1995bk}  & $0.3$ & $0$  \\[4pt] 
 Ananthanarayan \cite{Ananthanarayan:1998hj}  & $0.3$ & $0$  \\[4pt] 
 this work & $0.3$ & $0$ \\[4pt]  
  \hline
\end{tabular}
\end{center}
\caption{Tensor contributions to the $SU(2)$ LECs $l_1, l_2$ from
  various sources.}
\label{tab:LECfromLit}
\end{table}

\section{$\mathbf{1^{+-}}$ resonances} 
\label{sec:vector}
\renewcommand{\theequation}{\arabic{section}.\arabic{equation}}
\setcounter{equation}{0}
The contributions of axial-vector mesons with odd C-parity
($J^{PC}=1^{+-}$) to the LECs of $O(p^4)$ have not been considered up
to now. This may partly be due to the fact that the corresponding nonet
has not been unambiguously identified yet
\cite{Yao:2006px}. Only the states $h_1(1170)$ and $b_1(1235)$ are
listed in the PDG booklet as respectable resonances. Moreover, there
is only limited information on decay widths and branching ratios.
Nevertheless, there are good arguments for the existence of a complete
nonet (e.g., Ref.~\cite{Cirigliano:2003yq}). 
Although the masses of this nonet are considerably larger than those
of the lowest-lying vector mesons, they are comparable with the masses
of the axial-vector mesons with positive C-parity ($J^{PC}=1^{++}$). 
Since the latter make an important contribution to $L_{10}$ 
\cite{Ecker:1988te}, there is a priori no reason to disregard the
$1^{+-}$ nonet.

To investigate contributions of spin-1 exchange to the LECs
of $O(p^4)$, it is convenient to describe those mesons in terms of
antisymmetric tensor fields (see App.~\ref{app:vector}). 
Denoting the nonet spin-1 field as $H_{\mu\nu}$, the kinetic
Lagrangian is given by
\begin{equation}\label{eq:Hkin}
\mathcal L_H = \left\langle - \frac{1}{2}\nabla^\mu H_{\mu\nu} 
\nabla_\rho H^{\rho\nu} + \frac{M_H^2}{4} H_{\mu\nu} H^{\mu\nu} 
\right\rangle ~.
\end{equation}
As in the case of $1^{--}$ and $1^{++}$ exchange \cite{Ecker:1988te},
the $SU(3)$ singlet cannot contribute at $O(p^4)$.
Under parity and charge conjugation, the relevant octet field
$H_{\mu\nu}$ transforms as 
\begin{eqnarray}
	H_{\mu\nu}(t,\vec{x}) & \stackrel{P}{\rightarrow} & - 
\epsilon(\mu) \epsilon(\nu) H_{\mu\nu}(t,-\vec{x}) \nonumber \\
	H_{\mu\nu}(x) & \stackrel{C}{\rightarrow} & - H^T_{\mu\nu}(x).
\end{eqnarray}
The most general chiral invariant interaction of $O(p^2)$ of the 
$1^{+-}$ mesons with the Goldstone bosons respecting P and C
invariance is then given by
\begin{equation}\label{eq:LintH}
\mathcal L_{\rm int}[H(1^{+-})] = \left\langle H_{\mu\nu} J_H^{\mu\nu} 
\right\rangle
\end{equation}
with an antisymmetric tensor current
\begin{equation}
J_H^{\mu\nu} = \displaystyle\frac{F_{H}}{4\sqrt{2}}
~\ve^{\mu\nu\rho\sigma} f_{+ \,\rho\sigma} + \displaystyle\frac{i
  G_{H}}{2\sqrt{2}} ~\ve^{\mu\nu\rho\sigma} u_{\rho} u_{\sigma}.
\label{eq:Hcurr}
\end{equation}
As already pointed out, the $SU(3)$ singlet field does not couple
because of
\begin{equation}
	\left\langle J_H^{\mu\nu} \right\rangle = 0.
\end{equation}
Expanding around the classical solution in the usual manner, we 
obtain the effective action induced by $1^{+-}$ exchange 
\begin{equation}
S_H^{\rm eff} = \frac{1}{2} \int d^4 x \left\langle H^{\rm cl}_{\mu\nu} 
J_H^{\mu\nu} \right\rangle ~.
\end{equation}
To $O(p^4)$, the effective action is 
\begin{equation}
S_H^{\rm eff} = \int d^4 x \;\mathcal L^H_{4,{\rm bare}}(x)~,
\end{equation}
with the Lagrangian $\mathcal L^H_{4,{\rm bare}}$ given by
\begin{equation}\label{l4H}
\mathcal L^H_{4,{\rm bare}} = - \frac{1}{M^2_H} \left\langle
J_{H \,\mu\nu} J_H^{\mu\nu} \right\rangle ~.
\end{equation}
A straightforward calculation produces an effective Lagrangian
of the form (\ref{eq:L4}) with
\begin{eqnarray} 
& L^H_{1,{\rm bare}} = - \displaystyle\frac {G^2_H}{8 M_H^2}~, &
L^H_{2,{\rm bare}} = 2 L^H_{1,{\rm bare}}~, \quad
L^H_{3,{\rm bare}} = -6 L^H_{1,{\rm bare}}~, \nn
& L^H_{9,{\rm bare}} = - \displaystyle\frac {F_H G_H}{2 M_H^2}~, &
L^H_{10,{\rm bare}} = \displaystyle\frac {F^2_H}{4 M_H^2} ~. 
\label{eq:LHbare}
\end{eqnarray}
We have chosen the normalization of couplings in the current
(\ref{eq:Hcurr}) to facilitate comparison with vector meson
exchange. Comparing with the results of Ref.~\cite{Ecker:1988te}, one
finds that the replacements $F_V \to F_H$ and $G_V \to G_H$ yield the
LECs in Eq.~(\ref{eq:LHbare}) except for an overall change of
sign. Except possibly for $L_9$ (we do not know the relative sign of
$F_H,G_H$ in contrast to $F_V G_V > 0$), these results seem to suggest
that exchange of $1^{+-}$ resonances reduces the effect of vector meson
exchange. The relevant question is then: by how much?
 
The bare LECs $L^H_{i,{\rm bare}}$ ($i=1,2,3$) contribute to elastic
meson-meson scattering. In analogy to the case of vector mesons 
\cite{Ecker:1988te}, we are led to determine the coupling constant
$G_H$ from the decays of $1^{+-}$ resonances to two pseudoscalar
mesons. But parity conservation does not allow for such
decays. Consequently, $H$ exchange can only lead to a polynomial
contribution to the elastic meson-meson amplitude. For example, the
pion-pion scattering amplitude is given by
\begin{equation}\label{eq:AstuH}
A_H(s,t,u) = \frac{G_H^2}{M_H^2 F^4} (2 s^2 - t^2 - u^2),
\end{equation}
in accordance with (\ref{eq:LHbare}). But this form of the amplitude
is not compatible with the structure of the dispersion relations
discussed in Sec.~\ref{sec:SDtensor}. Therefore, the short-distance
constraints require the introduction of an additional contribution
from the $O(p^4)$ Lagrangian (\ref{eq:L4}) that completely cancels the
$H$ exchange contribution (\ref{eq:AstuH}):
\begin{equation}
A_{\rm SD}(s,t,u) = - A_H(s,t,u) ~. 
\end{equation}

Similarly, $H$ exchange contributes a term linear in $t$ to the vector 
form factor of the pion:
\begin{equation}
F_V^H (t) = - \displaystyle\frac{F_H G_H}{M_H^2 F^2}~t~.
\label{eq:FVt}  
\end{equation}
Again, there is no pole term because $H$ mesons cannot decay into two
pions. As discussed in Sec.~\ref{sec:resex}, the
absence of a pole contribution implies that
$H$ exchange does not contribute to $L_9$. 

Finally, we turn to the $VV - AA$ two-point function
\begin{eqnarray} 
i \displaystyle\int d^4x \,e^{\displaystyle{ipx}} 
\langle 0|T\left[V^i_\mu(x) V^j_\nu(0) - A^i_\mu(x) A^j_\nu(0) 
\right]|0\rangle & & \\
&  & \mbox{} \hspace*{-5cm} = \delta_{ij} \left[\left(p_\mu p_\nu - 
g_{\mu\nu}p^2
  \right) \Pi^{(1)}_{\rm LR} (p^2) + p_\mu p_\nu  \Pi^{(0)}_{\rm LR}
  (p^2)  \right] ~. \no
\end{eqnarray} 
According to QCD the invariant function $\Pi^{(1)}_{\rm LR} (p^2)$
satisfies an unsubtracted dispersion relation
\cite{Floratos:1978jb}. Again, 
$H$ exchange is incompatible with the short-distance constraint
because it produces a constant contribution corresponding to
$L^H_{10,{\rm bare}}$ in (\ref{eq:LHbare}). This contribution must
again be cancelled by a local counterterm leading to the final
conclusion that there are no $1^{+-}$ exchange contributions to the
LECs of $O(p^4)$ at all:
\begin{equation}
L^H_i = 0 \qquad (i=1,\dots,10) ~.
\end{equation}

\section{Conclusions} 
\label{sec:concl}
The saturation of low-energy constants of $O(p^4)$ by the exchange of
$V,A,S$ and $P$ meson resonances is a generally accepted
feature of strong dynamics at low energies. Chiral vector meson
dominance can easily be understood because of the strong coupling to
the pseudoscalars and the comparatively low masses of the lowest-lying
vector meson nonet. On the other hand, it is much less obvious why
$A,S$ and $P$ resonances should be more important than the $2^{++}$
and $1^{+-}$ states with similar masses. The latter two multiplets
complete the spectrum of $\overline{q} q$ bound states with orbital 
angular momentum $\le 1$. 

Setting up the most general chiral resonance Lagrangians for $2^{++}$ and
$1^{+-}$ fields and integrating out the resonance fields, the tensor
meson contributions to the LECs $L_1$, $L_2$ and $L_3$ seem to depend
on a coupling that cannot be determined from tensor meson decays. In
the case of $1^{+-}$ exchange, the same LECs are affected that
receive vector meson contributions, albeit with opposite sign. Both
results are superficial and must be confronted with the short-distance
constraints of QCD.

In the tensor meson case, the constraints of
axiomatic field theory for elastic meson-meson scattering are actually
sufficient to show that only $L_3$ receives a non-zero
contribution. The resulting value $L_3^T = 0.16 \cdot 10^{-3}$ is
completely negligible compared to the sum of vector and scalar 
contributions. Our results agree with those in 
Refs.~\cite{Toublan:1995bk,Ananthanarayan:1998hj} in the limit of
chiral $SU(2)$ but
we disagree with all other predictions in the literature. 

The final results for $1^{+-}$ exchange are even more pronounced. The
combined short-distance constraints for elastic meson-meson
scattering, the vector form factor of the pion and the $VV -AA$
two-point function eliminate all contributions of $1^{+-}$ exchange to
the LECs of $O(p^4)$.

The final conclusion can be summarized in one sentence: the
dominance of $V,A,S,P$ exchange contributions to the LECs of $O(p^4)$
is not an accident.

\section*{Acknowledgements}
We are particularly grateful to J\"urg Gasser for many helpful
discussions and for providing us with his notes on symmetric tensor
fields. We also thank Balasubramanian Ananthanarayan, Roland Kaiser, 
Heiri Leutwyler, Helmut Neufeld, Jorge Portol\'es and Matthew Schwartz 
for clarifying comments.

\appendix
\renewcommand{\theequation}{\Alph{section}.\arabic{equation}}
\renewcommand{\thetable}{\Alph{section}.\arabic{table}}
\setcounter{equation}{0}
\setcounter{table}{0}

\section{Symmetric tensor fields for spin 2}
\label{app:tensor}

The Lagrangian for a hermitian spin-2 field $T_{\mu\nu}$ coupled
linearly to a source $J_{\mu\nu}$ can be written in the
form \cite{rivers,Bellucci:1994eb}
\begin{eqnarray}
	\mathcal L = - \frac{1}{2} T_{\mu\nu} D^{\mu\nu,\rho\sigma}_T 
T_{\rho\sigma} + T_{\mu\nu} J^{\mu\nu} ~,
\label{eq:L_T}
\end{eqnarray}
with $T_{\mu\nu}=T_{\nu\mu}$, $J_{\mu\nu}=J_{\nu\mu}$ and 
\begin{eqnarray}
D^{\mu\nu,\rho\sigma}_T & = & (\DAl + M_T^2 )\left[ 
\frac{1}{2}(g^{\mu\rho}g^{\nu\sigma} + 
g^{\mu\sigma}g^{\nu\rho})-g^{\mu\nu}g^{\rho\sigma} \right] 
\nonumber \\ && 
+ g^{\rho\sigma} \partial^\mu \partial^\nu +g^{\mu\nu} 
\partial^\rho \partial^\sigma -
\frac{1}{2}(g^{\nu\sigma} \partial^\mu \partial^\rho +
g^{\rho\nu} \partial^\mu \partial^\sigma + g^{\mu\sigma} 
\partial^\rho \partial^\nu + g^{\rho\mu} \partial^\sigma
\partial^\nu).
\label{eq:D_T} 
\end{eqnarray}
The Feynman propagator is given by
\begin{eqnarray}\label{eq:tensorprop}
	G^T_{\mu\nu,\rho\sigma}(x) &=&  \displaystyle\int  
\displaystyle\frac{d^4 k}{(2\pi)^4} \displaystyle\frac{e^{-ikx} 
P_{\mu\nu,\rho\sigma}(k)}{M_T^2-k^2-i\epsilon} 
\nonumber \\
P_{\mu\nu,\rho\sigma} &=& \frac{1}{2}
(P_{\mu\rho}P_{\nu\sigma}+P_{\nu\rho}P_{\mu\sigma})- 
\frac{1}{3} P_{\mu\nu}P_{\rho\sigma} \\   
P_{\mu\nu}& = &g^{\mu\nu} - \frac{k_\mu k_\nu}{M_T^2}~,\nonumber
\end{eqnarray}
satisfying the differential equation
\begin{eqnarray} 
D^{\mu\nu,\lambda\rho}_T G^T_{\lambda\rho,\sigma\tau}(x) &=&
\displaystyle\frac{1}{2} \left(\delta^\mu_\sigma \delta^\nu_\tau +
\delta^\nu_\sigma \delta^\mu_\tau \right) \delta^{(4)} (x)~.
\end{eqnarray}
The classical equation of motion
\begin{equation}
D^{\mu\nu,\rho\sigma}_T T_{\rho\sigma} = J^{\mu\nu}
\label{eq:EOM_T}
\end{equation}
has the solution
\begin{equation}
T_{\mu\nu}^{\rm cl}(x) = \displaystyle\int d^4 y\,  
G^T_{\mu\nu,\rho\sigma}(x-y) J^{\rho\sigma}(y) ~.
\end{equation}
Without the inclusion of auxiliary fields in the Lagrangian 
\cite{Fierz:1939ix,Toublan:1995bk},
the tensor field $T_{\mu\nu}$ is neither traceless nor
transverse. However, the corresponding components do not propagate in
accordance with the spin-2 nature of the field:
\begin{eqnarray}
P^\mu_{~\mu,\rho\sigma}(k) &=& \displaystyle\frac{k^2 - M_T^2}{3 M_T^2}
\left(g_{\rho\sigma} + \displaystyle\frac{2 k_\rho k_\sigma}{M_T^2} 
\right) \\
k^\mu P_{\mu\nu,\rho\sigma}(k) &=& \displaystyle\frac{M_T^2 - k^2}{6
  M_T^2} \left(3 k_\rho P_{\nu\sigma} + 3 k_\sigma P_{\nu\rho} - 2
k_\nu P_{\rho\sigma} \right) ~.\no
\end{eqnarray}
The one-particle matrix element for a spin-2 particle with momentum $k$
and polarization $\lambda$ is expressed in terms of the polarization
tensor $\ve_{\mu\nu} (k;\lambda)$:
\begin{eqnarray}
\label{eq:ME_T} 
\left\langle 0 \left| T_{\mu\nu}(0) \right| T(k;\lambda) 
\right\rangle &=& \ve_{\mu\nu} (k;\lambda) \\
\ve_{\mu\nu} = \ve_{\nu\mu}~, && k^\mu \ve_{\mu\nu} =
0~, \qquad \ve^\mu_{~\mu} = 0 ~. \nonumber
\end{eqnarray}
The explicit form of the polarization tensor can be found, e.g., in
Ref.~\cite{Pilkuhn:1979ps}. For the decay rate of an unpolarized
spin-2 particle one needs the sum over polarizations
\begin{equation}
\displaystyle\sum_\lambda \ve_{\mu\nu} (k;\lambda)
\ve_{\rho\sigma} (k;\lambda)^* = P_{\mu\nu,\rho\sigma}(k) 
\end{equation}
where $P_{\mu\nu,\rho\sigma}(k)$ is defined in Eq.~(\ref{eq:tensorprop}).

\setcounter{equation}{0}
\setcounter{table}{0}

\section{Antisymmetric tensor fields for spin 1}
\label{app:vector}
For completeness, we collect in this appendix a few basic formulas for
the description of spin-1 fields in terms of antisymmetric tensor
fields. 

The Lagrangian for a hermitian spin-1 field $H_{\mu\nu}$ coupled
linearly to a source $J_{\mu\nu}$ can be written in the
form (e.g., App. A in Ref.~\cite{Ecker:1988te})
\begin{eqnarray}
	\mathcal L = \frac{1}{2} H_{\mu\nu} D^{\mu\nu,\rho\sigma}_H 
H_{\rho\sigma} + H_{\mu\nu} J^{\mu\nu} ~,
\end{eqnarray}
with $H_{\mu\nu}= - H_{\nu\mu}$, $J_{\mu\nu}= - J_{\nu\mu}$ and 
\begin{eqnarray}
D^{\mu\nu,\rho\sigma}_H & = & \displaystyle\frac{1}{4} ~\partial_\lambda 
\left[g^{\rho\lambda} \left(\partial^\mu g^{\nu\sigma} - \partial^\nu
  g^{\mu\sigma} \right) - g^{\sigma\lambda} \left(\partial^\mu
  g^{\nu\rho}  - \partial^\nu g^{\mu\rho} \right) \right] \nn  
 && + \frac{M_H^2}{4}\left(g^{\mu\rho}g^{\nu\sigma} - 
g^{\mu\sigma}g^{\nu\rho} \right) ~.
\end{eqnarray}
The Feynman propagator is given by
\begin{eqnarray}\label{eq:vectorprop}
	G^H_{\mu\nu,\rho\sigma}(x) &=&  \displaystyle\int  
\displaystyle\frac{d^4 k}{(2\pi)^4} \displaystyle\frac{e^{-ikx} 
Q_{\mu\nu,\rho\sigma}(k)}{M_H^2 (M_H^2-k^2-i\epsilon)} 
\nonumber \\
Q_{\mu\nu,\rho\sigma} &=& \left[g_{\mu\rho} g_{\nu\sigma} \left(M_H^2-
  k^2 \right) + g_{\mu\rho} k_\nu k_\sigma - g_{\mu\sigma} k_\nu
  k_\rho - \left(\mu \leftrightarrow \nu  \right) \right]  
\end{eqnarray}
satisfying the differential equation
\begin{eqnarray} 
D^{\mu\nu,\lambda\rho}_H G^H_{\lambda\rho,\sigma\tau}(x) &=&
\displaystyle\frac{1}{2} \left(\delta^\mu_\sigma \delta^\nu_\tau -
\delta^\nu_\sigma \delta^\mu_\tau \right) \delta^{(4)} (x)~.
\end{eqnarray}
The classical equation of motion
\begin{equation}
D^{\mu\nu,\rho\sigma}_H H_{\rho\sigma} = - J^{\mu\nu}
\end{equation}
has the solution
\begin{equation}
H_{\mu\nu}^{\rm cl}(x) = - \displaystyle\int d^4 y\,  
G^H_{\mu\nu,\rho\sigma}(x-y) J^{\rho\sigma}(y) ~.
\end{equation}
The one-particle matrix element for a spin-1 particle with momentum $k$
and polarization $\lambda$ is expressed in terms of the usual
polarization vector $\ve_{\mu} (k;\lambda)$:
\begin{eqnarray} 
\left\langle 0 \left| H_{\mu\nu}(0) \right| H(k;\lambda) 
\right\rangle &=& i M_H^{-1} \left[k_\mu \ve_{\nu} (k;\lambda)-
 k_\nu \ve_{\mu} (k;\lambda) \right]~, \qquad 
 k^\mu \ve_{\mu} = 0 ~.
\end{eqnarray}


\end{document}